\documentclass[epj]{svjour}

\usepackage{epsfig}
\usepackage{amsmath}
\usepackage{amssymb}
\usepackage{bbold}

\newcommand{\bra}[1]{\mbox{$\langle #1|$}}
\newcommand{\ket}[1]{\mbox{$|#1\rangle$}}

\newcommand{\eqn}[1]{\begin{equation} #1 \end{equation}}
\newcommand{\eqna}[1]{\begin{eqnarray} #1 \end{eqnarray}}
\newcommand{\dstyle}[1]{$\ensuremath{\displaystyle{#1}}$}
\newcommand{\rmd}{{\ensuremath{\rm d}}}

\newcommand{\bs}[1]{\ensuremath{\boldsymbol{#1}}}

\renewcommand{\Im}{{\ensuremath{\rm Im}}}

\begin{document}

\title{
%\hfill{\small {\bf MKPH-T-05-07}}\\
  $\pi^0$-Photoproduction on the deuteron via $\Delta$-excitation
  using the Lorentz Integral Transform 
}

\author{
  Christoph Rei{\ss}\inst{1,2}, 
  Hartmuth Arenh{\"o}vel\inst{1}
  and
  Michael Schwamb\inst{1}
}

\institute{
Institut f\"ur Kernphysik, Johannes Gutenberg-Universit\"at
  Mainz, D-55099 Mainz, Germany
  \and
Dipartimento di Fisica, Universit\`a di Trento, I-38050 Povo, Italy
}

\date{\today}

\abstract{
  The Lorentz Integral Transform method (LIT) is extended to  
  pion photoproduction in the $\Delta$-resonance region. The main focus 
  lies on the solution of the conceptual difficulties which arise if 
  energy dependent operators for nucleon resonance excitations are considered.
  In order to demonstrate the applicability of our approach, we calculate
  the inclusive cross section for $\pi^0$-photoproduction off the deuteron 
  within a simple pure resonance model.
}

\PACS{
  {02.30.Uu}{Integral transforms --}
  {13.60.Le}{Meson production --}
  {21.45.+v}{Few-body systems --}
  {25.10.+s}{Nuclear reactions involving 
    few-nucleon systems --}
  {25.20.Lj}{Photoproduction reactions}
  }

\titlerunning{
  $\pi^0$-Photoproduction on  the deuteron via $\Delta$-excitation
  using the LIT method
}

\authorrunning{
  Christoph Rei{\ss} {\it et al.}
}

\date{\today}

\maketitle

\section{Introduction}
\label{sec:intro}

The Lorentz Integral Transform method (LIT)~\cite{elo1} has been proven
to be a powerful technique for calculating inclusive
(see {\it e.g.}~\cite{mkog,PRL3,Sonia} and references therein)
%\cite{elo1,elo2,elo3a,elo3b,elo3c,elo4,elo5a,elo5b,elo6,mkog,PRL3,Sonia}
as well as exclusive~\cite{efros85,lapwl,Sofia} photoreaction cross sections 
with complete inclusion of final state interaction (FSI) without 
calculating final continuum wave functions. Recently, the technique has 
also been extended to electroweak processes~\cite{Nir}.

This success has motivated the extension of the LIT method to pionproduction
processes on light nuclei $A \geq 3$, where existing approaches
\cite{Lag81,KaT95} call for considerable improvements.
In the case of $^3$He, for example, a conventional treatment of FSI would 
imply a four-body Faddeev-Yaku\-bov\-sky treatment~\cite{Glo83} of the 
final state which is very complicated. With respect to present 
experimental programs to study electromagnetic meson production on 
light nuclei, {\it e.g.}~at MAMI in Mainz, more sophisticated 
calculations of such reactions are certainly needed in the near future.
 
Recently, the LIT has been applied to inclusive pion photoproduction
on the deuteron as the simplest possible nuclear target~\cite{Christoph}.
As a first step, only the near threshold region has been considered,
where only the dominant Kroll-Ruderman~\cite{KR} as production 
operator was included, and results comparable to traditional approaches 
were a\-chiev\-ed. 

In the present work we want to extend this approach to higher photon 
energies into the $\Delta(1232)$-resonance region. As we will see below,  
one cannot proceed naively in a straightforward manner, because of
conceptual problems which arise from the energy dependence associated with 
the resonance contribution to the elementary production operator. 
We will first give in section~\ref{sec:lit} a brief outline
of the LIT approach for energy independent transition operators. 
The problem of the standard LIT method for electromagnetic 
particle production via resonances is discussed in section~\ref{sec:litp},
where we also present a formal solution. As a test case, we consider in 
section~\ref{sec:litpi0} a simple model for $\pi^0$-production on the 
deuteron in the $\Delta$-region. The corresponding results,
together with a summary and an outlook, are presented in 
section~\ref{sec:res}. 

\section{
  The Lorentz Integral Transform method for energy independent
  transition operators}\label{sec:lit}

In this section we review briefly the LIT method 
for inclusive reactions. The central quantity is the response function 
\begin{equation}
  \label{Resp}
  R(\omega)
  =\int\!\rmd\Psi_f\,
  |\langle\Psi_f| O|\Psi\rangle|^2 
  \delta(E_f-E_0-\omega) 
  \,,
\end{equation}
where $O$ is an operator describing the transition from the ground 
state $\Psi$, with energy $E_0$, to final states 
$\Psi_f$, with energies $E_f$, in the specific process 
under consideration. In the LIT approach, the response function
is not calculated directly. Rather, one first introduces an integral 
transform of the response function $R(\omega)$ by
\eqn{\label{eqLIT}
  L(\sigma) = \int_{\omega_{\rm th}}^\infty\!\rmd\omega\,
  \frac{R(\omega)} {(E_0+\omega-\sigma_{\rm R})^2 + \sigma_{\rm I}^2} 
  \,,}
where $\sigma=\sigma_{\rm R}+i\,\sigma_{\rm I}$ with $\sigma_{\rm I} \not= 0$. 
Furthermore, ${\omega_{\rm th}}$ denotes the reaction threshold.

Inserting the response function from eq.~(\ref{Resp}), one can rewrite
 eq.~(\ref{eqLIT}) 
 as follows 
\eqna{\label{eqn:lit1}
  L(\sigma)
  & = & 
  \int\!\rmd W\rmd \Psi_f\,\delta(E_f-W)\,
  \nonumber\\
  &&
  \bra{\Psi}\,O^\dagger\,
  \frac{1}{W-\sigma^*}\ket{\Psi_f} \bra{\Psi_f}\frac{1}{W-\sigma}
  \,O\,\ket{\Psi}\,,
  \nonumber\\
  && \hspace*{-.9cm}= 
  \int\!\rmd \Psi_f\,
  \bra{\Psi}\,O^\dagger\,
  \frac{1}{E_f-\sigma^*}\ket{\Psi_f} \bra{\Psi_f}\frac{1}{E_f-\sigma}
  \,O\,\ket{\Psi}
  \,.
}
Now the Schr{\"o}dinger equation
$H\ket{\Psi_f}=E_f\ket{\Psi_f}$ can be used to replace $E_f$ with the
hamiltonian $H$ of the given system
\eqn{\label{eqn:comp1}
  L(\sigma)
  =
  \int\!\rmd \Psi_f
  \,
  \bra{\Psi}\,O^\dagger\,
  \frac{1}
  {H-\sigma^*}
  \ket{\Psi_f}\bra{\Psi_f}
  \frac{1}
  {H-\sigma}
  \,O\,\ket{\Psi} \,\, .
}
In this step it is essential that the operator $O$ is energy
 independent. By using the completeness relation
\eqn{\int\!\rmd \Psi_f\,\ket{\Psi_f}\bra{\Psi_f}=\mathbb 1\,,} 
one obtains finally 
\eqna{\label{eqn:comp2}
  L(\sigma)
  & = &
    \bra{\Psi}\,O^\dagger\,
    (H-\sigma^*)^{-1} \,
    (H-\sigma)^{-1}
    \,O\,\ket{\Psi}\nonumber \\ 
    &=&\langle\,\widetilde{\Psi}\,|\,\widetilde{\Psi}\,\rangle\,,
}
where one has introduced the so-called Lorentz state 
\eqn{|\,\widetilde{\Psi}\,\rangle=(H-\sigma)^{-1}\,O\,\ket{\Psi}\,,}
which obeys an inhomogeneous differential equation 
\eqn{\label{eqn:lit}
  \left({H-\sigma}\right)\ket{\,\widetilde{\Psi}}= \,O\,\ket{\Psi}\,\, ,
}
and which is bound at infinity.
Recalling the fact that $H$ is hermitean and therefore has only real
eigenvalues, this feature guarantees a unique solution of (\ref{eqn:lit})
because the corresponding homogeneous equation has
only the trivial solution. Since the source at the right hand side of
(\ref{eqn:lit}) is localized and $\Im\{\sigma\}\not=0$, the asymptotic
behaviour  of $\widetilde{\Psi}$  at infinity is bound-state like. 
Thus the evaluation of $L(\sigma)$ avoids the explicit calculation of 
the continuum states $\Psi_f$ but still includes the complete final
state interaction. In the final step, the desired response function $R$
is obtained from $L(\sigma)$ by an appropriate inversion method,
see~\cite{diego} concerning further details. 

\section{The Lorentz Integral Transform method for meson production
processes}\label{sec:litp}

In the foregoing derivation an essential assumption was that the
transition operator $O$ is energy independent. This is, for example,
the case for dipole absorption in the long wave length limit, but it
is certainly no longer fulfilled for a retarded dipole operator. The
same is true for the Kroll-Ruderman term in~\cite{Christoph}. However,
in both cases the energy dependence is smooth and weak. In principle,
one could be tempted to replace then in the operator the energy by the
corresponding hamiltonian. But then the equation for the Lorentz state
becomes quite complicated and non-linear in the hamiltonian. In order
to avoid this complication, another method has been devised~\cite{Christoph} by
treating the energy in the operator as a parameter, fixed to some
value $\epsilon$. One then determines a Lorentz transform $\widetilde
L(\sigma,\epsilon)$ as a function of this additional variable
$\epsilon$. The inversion then yields a response 
function $\widetilde R(\omega,\epsilon)$ from which the desired
response function is obtained by setting
\eqn{R(\omega)=\widetilde R(\omega,\omega)\,.}

\begin{figure}[ht]
\centerline{\epsfig{file=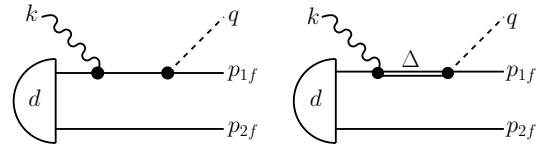,width=7cm}}
\caption{\label{fig:01} Nucleon- and $\Delta$-pole contributions to pion 
photoproduction on the deuteron. }
\end{figure}

However, this method fails for a resonance-like energy dependence which
usually appears when certain degrees of freedom or states are
projected out in 
favor of effective operators. This is the case in pion photoproduction
where any realistic production operator contains resonance or pole
contributions (see for example~\cite{blomqvist,DrH99}) describing
intermediate excitation of nucleon isobars. The problem, one
encounters, can be illustrated already by the contribution of the
nucleon-pole diagram to the production process on the two nucleon
system depicted in Fig.~\ref{fig:01}, which is an essential ingredient in
pion production. Projecting out the intermediate $NN$-system in time
ordered perturbation theory, the corresponding amplitude has the
following structure 
\eqn{\label{eqn:pol}
  O(E_f) \propto \frac{1}{E_f+i\,\varepsilon-H_0}\,,
  \,
}
where $H_0$ denotes the free $NN$-hamiltonian. 
Recalling now the different steps in equation (\ref{eqn:comp1}) and
(\ref{eqn:comp2}), and applying the LIT approach naively would mean to
replace the energy $E_f$ appearing in the propagator by the full
hamiltonian $H$ of the final $\pi NN$-system. This, however, does not
make any sense, because $H_0$ and $H$ act on different particle
systems. Also the method of treating the energy as a parameter does
not work, because in this case the effective operator becomes singular
and thus the Lorentz state is not normalizable any more.

The physical reason for the energy dependence of the contribution
(\ref{eqn:pol}) is quite obvious because the intermediate
$NN$-configuration has been projected out in favor of an effective
operator. The same situation appears for the $\Delta$-resonance
contribution in Fig.~\ref{fig:01}, 
which is also described by an effective, energy
dependent operator. A way out of this dilemma is to enlarge the
configuration space by including the previously projected-out states
and thus avoiding the energy dependence of effective operators. 

In the present case, where a meson is produced, we have to include
configurations with different numbers of particles, {\it i.e.}~we have to
construct a genuine Fock space description. We will first consider a
general Fock space ${\cal F}$ consisting of $N$ orthogonal subspaces
labeled by $l$. The projectors $P_l$ onto these subspaces fulfil
\eqna{
  \mathbb 1 & = & \sum_{l=1}^N P_l\, \quad \mbox{with} \quad  
  P_l P_m =  \delta_{lm}P_l\,.
}
The full Fock space hamiltonian $H=H_0+V$ consists of a kinetic part
$H_0$ which is diagonal and an interaction $V_{lm} = P_l V P_m$, with
$l,m\in\{1,...,N\}$, allowing transitions between the various
subspaces. A transition operator, describing an electromagnetic
process, is generally expressed by a hermitean current operator
$j^{\mu}$ in ${\cal F}$, which does not contain any resonance-like
energy dependence,
\eqn{\label{eqn:current1}
  O =\sum_{l,m=1}^N \varepsilon_\mu j^\mu_{lm} \,\, \quad
 \mbox{with} \quad  j^\mu_{lm} = P_l j^\mu P_m\,\,,
}
where $\varepsilon^\mu$ denotes the photon polarization vector.
A remaining trivial energy dependence via current structures depending
on the photon momentum can be handled by the parameter method. 

In this Fock space, the LIT equation reads now 
\eqn{
  (H-\sigma)\ket{\widetilde{\Psi}} = O \ket{\Psi}
  \,,\,\Im\{\sigma\}\not=0\,.
}
To solve this equation we use the method of resolvents 
in analogy to standard scattering theory, {\it i.e.}~we introduce
\eqna{
  {\cal G}_0(\sigma)  &=&  \frac{1}{\sigma-H_0}\,,\quad
  {\cal G}(\sigma)    =  \frac{1}{\sigma-H}\,.
}
Defining a conventional $T$-matrix by $T{\cal G}_0 \equiv V{\cal G}$,
one then obtains the following standard equations for the ${\cal G}$- and
$T$-matrices
\eqna{
  {\cal G} 
  & = & {\cal G}_0 + {\cal G}_0 T {\cal G}_0
   =  {\cal G}_0 + {\cal G}_0 V {\cal G}\,,\\% \quad \mbox{with} \quad 
  T  &=&  V + V{\cal G}_0 T\,. 
}
A formal solution of the LIT-equation can be written as
\eqn{
  \ket{\widetilde{\Psi}} = -{\cal G}O\ket{\Psi}
  \,.
}
Since $O$ is regular and $\Im\{\sigma\}\not=0$ , the Lorentz state
$\widetilde{\Psi}$ is normalizable like a bound-state.

\section{Application to $\pi^0$-photoproduction via the excitation of
the $\Delta$-resonance}\label{sec:litpi0} 

\begin{figure}[ht]
\centerline{\epsfig{file=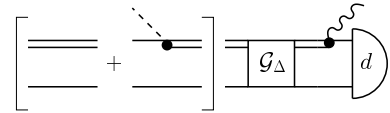,width=5cm}}
\caption{\label{fig:02}
  Diagrammatic representation of $\widetilde{\Psi}$ in eq.~(\ref{psitilde}).}
\end{figure}

As a test case, we will consider now $\pi^0$-photoproduction off the
deuteron in the $\Delta$-region. Then ${\cal F}$ comprises besides
$\pi^0 NN$- also  $NN$- and $N\Delta$-states, so that nucleon-pole as
well as $\Delta$-pole contributions to the photoproduction process can
be taken into account. Using a compact matrix notation with respect to
the three projectors $P_\pi$, $P_N$ and $P_\Delta$ for the $\pi^0 NN$,
$NN$- and $N\Delta$-subspaces, respectively, we can cast kinetic and
interaction parts of the hamiltonian $H=T+V$ into the following form 
\eqn{\label{eqn:v}
  H_0 = 
  \left(
    \begin{matrix}
      T_{N} \\ 
      & T_{\Delta } \\ 
      && T_{\pi}
    \end{matrix}
  \right)\, \, , 
 \quad 
V=  \left(
    \begin{matrix}
      V_{NN}  & V_{N \Delta } & V_{N \pi } \\ 
      V_{\Delta N }  & V_{\Delta\Delta  } & V_{\Delta \pi } \\
      V_{\pi N }  & V_{\pi \Delta } & V_{\pi\pi } 
    \end{matrix}
  \right) \,\, . 
}
Transitions between different subspaces are described by the
non-diagonal matrix elements of $V$. For example, $V_{\pi N }$
contains the $\pi^0 N$-vertex allowing transitions between the $NN$-
and $\pi^0 NN$-states. Since we only want to demonstrate the
applicability of our approach and do not aim at a quantitative
description of the data, we restrict the interaction solely to
$V_{\Delta \pi}$ and $V_{NN}$, {\it i.e.}~we use 
\eqn{\label{eqn:v2}
  V = 
  \left(
    \begin{matrix}
      V_{NN}  & 0 & 0 \\ 
      0  & 0 & V_{\Delta \pi } \\
      0  & V_{\pi \Delta } & 0 
    \end{matrix}
  \right)
  \,.
}
The parametrization of the $\Delta N\pi^0$-vertex $V_{\Delta \pi}$ is taken
from~\cite{PoS87}. As electromagnetic current we consider here solely
the dominant M1-$N\Delta$-current $\bs{j}_{N\Delta}$ 
\eqn{\label{eqn:delta1}
\bs{j}_{\Delta N}(\bs{k}) 
  \sim
  \frac{G_{M1}^{0\,\Delta N}}{2M_N}
  \,i\,\bs{\sigma}_{\Delta N}\times\bs{k}\,\,
}
with \dstyle{G_{M1}^{0\,\Delta N}=4.22}. With these building blocks we
can describe the dominant $\Delta$-pole contribution to
photopionproduction. With the approximations (\ref{eqn:v2}) and 
(\ref{eqn:delta1}), the
$NN$-interaction $V_{NN}$ contributes solely to the deuteron
ground-state. For reasons of simplicity, we have used a pure $S$-wave
Yamaguchi potential~\cite{Yam} with modern parameters
from~\cite{Dress}. For a shorter notation we label operators $A$
connecting the same subspace with only one index, 
{\it i.e.}~$A_{ll}\equiv A_l$. 
The Lorentz state can then be written as (see Fig.~\ref{fig:02}) 
\eqna{\label{psitilde}
  -\ket{\widetilde{\Psi}}
  & = &
  {\cal G} O_{\Delta N} \ket{\Psi}  = 
  (\mathbb 1_{\Delta} + {\cal G}_{0\pi} V_{\pi\Delta}) {\cal G}_{\Delta}
  O_{\Delta N} \ket{\Psi}\,.
}
The resolvent ${\cal G}_{\Delta}$ fulfils the equation 
 (see Fig.~\ref{fig:03})
\eqna{\label{eqn:g0d}
  {\cal G}_{\Delta}
  & = & 
  {\cal G}_{0\Delta} 
  + 
  {\cal G}_{0\Delta} V_{\Delta \pi} {\cal G}_{0\pi} V_{\pi \Delta}
  {\cal G}_{\Delta} \,,
}
which allows one to rewrite the LIT into the following form
\eqn{\label{eq:flit}
  L 
  = 
  -\frac{1}{\sigma_{\rm I}}\,
  \Im\left\{ 
    \bra{\Psi} O_{N \Delta }\,{\cal G}_\Delta \, O_{\Delta N }\ket{\Psi}
  \right\}
  \,.
}
$V_{\Delta}$ consists of a loop diagram (disconnected) and a genuine
two-body one-pion exchange potential (connected), see
Fig.~\ref{fig:03}. If the latter is neglected, we will refer to it as
impulse approximation (IA) and denote the corresponding resolvent from
here on as ${\cal G}^{\rm IA}_{\Delta}$. Therefore in impulse
approximation $L^{\rm IA}$ is given by (\ref{eq:flit}) where ${\cal
G}_{\Delta}$ is replaced by
\eqn{
  {\cal G}^{\rm IA}_{\Delta}
  =
  \frac{1}{\sigma-T_{\Delta } - {\Sigma_{\Delta} (\sigma )}}\,,\label{impa}
}
with $\Sigma_{\Delta}$ as self-energy of the $\Delta$ 
\eqn{{
\Sigma_{\Delta} (\sigma) = 
V_{\Delta \pi}   {\cal G}_{0\pi}(\sigma)  V_{\pi \Delta} |_{\rm disconnected}
\,\, .
}}
 
\section{Results}
\label{sec:res}

\begin{figure}[ht]

\centerline{\epsfig{file=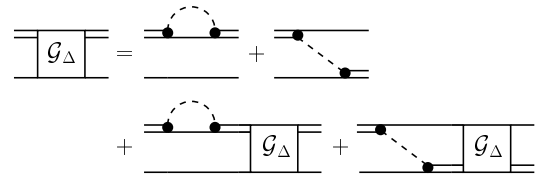,width=7cm}}
\caption{\label{fig:03}
  Diagrammatic representation of ${\cal G}_{\Delta}$ in eq.~(\ref{eqn:g0d}).
}
\end{figure}

\begin{figure}[ht]
\centerline{\epsfig{file=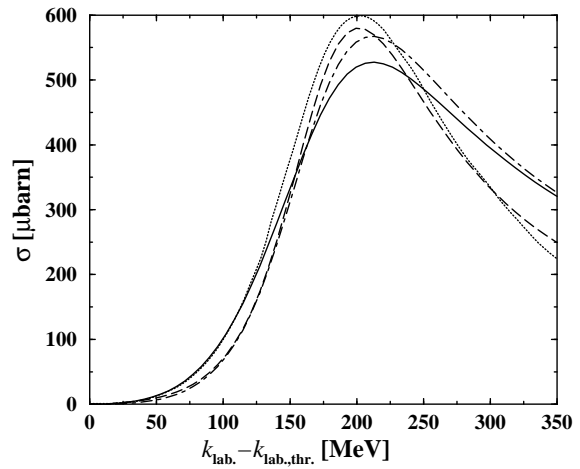,width=7.5cm}}
\caption{\label{fig:04}
  Total cross section of $\pi^0$-photoproduction on the deuteron. 
  Notation of the curves:
  dashed: IA with constant $\Delta$-mass;
  dashed-dotted: IA using eq.~(\ref{impa}); 
  full: inclusion of $N\Delta$-interaction; dotted: IA according 
  to~\cite{schmidt96} (only resonance). 
}
\end{figure}

The result for $\pi^0$-photoproduction on the deuteron
is shown in Fig.~\ref{fig:04}. Besides the IA according to (\ref{impa})
with a variable $\sigma$-dependent $\Delta$-mass,
we have calculated in addition another IA with a static $\Delta$-mass 
($M_\Delta=1232$\,MeV) for a better comparison with the 
IA of~\cite{schmidt96}. 
Considering the small model differences to~\cite{schmidt96} the 
agreement is quite satisfactory. While the IA is calculated without a
partial wave decomposition, it is needed for the
$N\Delta$-interaction contribution generated by the one-pion exchange diagram
in Fig.~\ref{fig:03}. In order to fulfil convergence, we took  into
account all channels with a total angular momentum $J \leq 5$ of the 
$N\Delta$-system. One readily notes that the $N\Delta$-interaction 
has a sizeable influence near threshold leading to an enhancement and 
is still moderate near the maximum resulting in a slight upward shift 
of the position and lowering of the absolute size by about 8~\%. 
For a realistic description of the cross section in the $\Delta$-region
one has to include besides non-resonant photoproduction contributions 
the neglected FSI, i.e.\ $NN$- and $\pi N$-interactions~\cite{DaA03}.

In conclusion, we have demonstrated how to extend the LIT method when 
energy dependent effective operators are involved. It turns out that
in such a case those states, which have been projected out in favor of
effective operators, have to be included explicitly in an expanded
Fock space. With respect to the specific process considered here as an
example, pion photoproduction on the deuteron, it is obvious that in
the future the present model has to be extended towards a more
realistic description of this reaction. First of  all, additional FSI
as well as
nonresonant Born contributions have to be included. Furthermore, since
this method really pays-off for more complex systems with more than
two nucleons, one has to consider meson-photoproduction on other light
nuclei like $^3$He or $^4$He.

This work was supported by the Deutsche Forschungsgemeinschaft (SFB 443).

%%% FIGURES %%%

\begin{thebibliography}{00}

\bibitem{elo1} V.D. Efros, W. Leidemann, and G. Orlandini,  
  Phys. Lett. {\bf B338} 130 (1994).

\bibitem{mkog} S. Martinelli, H. Kamada, G. Orlandini, and W. Gl\"ockle, 
  Phys. Rev. {\bf C 52} 1778 (1995).

\bibitem{PRL3} S. Bacca, M. Marchisio, N. Barnea, W. Leidemann, and
G. Orlandini, Phys. Rev Lett. {\bf 89}, 052502 (2002). 

\bibitem{Sonia} S. Bacca, H. Arenh\"ovel, N. Barnea, W. Leidemann, and
G. Orlandini, 
  Phys. Lett. {\bf B 603} 159 (2004). 

\bibitem{efros85} V.D. Efros, 
  Sov. J. Nucl. Phys. {\bf 41} 949 (1985).

\bibitem{lapwl} A. La Piana, and W. Leidemann, 
  Nucl. Phys. {\bf A677} 423 (2000). 

\bibitem{Sofia}  S. Quaglioni, W. Leidemann, G. Orlandini, N. Barnea, 
  and V.D. Efros, Phys. Rev. {\bf C 69}, 044002 (2004).

\bibitem{Nir} D. Gazit and N. Barnea, 
  Phys. Rev. {\bf C 70}, 048801 (2004).

\bibitem{Lag81} J. M. Laget, Phys. Rep. {\bf 69}, 1 (1981)

\bibitem{KaT95} S. S. Kamalov, L. Tiator, and C. Bennhold,
 Phys. Rev. Lett. {\bf 75} 1288 (1995). 

\bibitem{Glo83} W. Gl\"ockle, The Quantum Mechanical
 Few-Body Problem, Springer (1983). 

\bibitem{Christoph} C. Rei{\ss}, W. Leidemann, G. Orlandini, and 
  E.L. Tomusiak, Eur. Phys. J.  {\bf A 17}, 589 (2003).

\bibitem{KR} N.M. Kroll and M.A. Ruderman, 
  Phys. Rev. {\bf 93} 233 (1954).

\bibitem{diego} D. Andreasi, W. Leidemann, C. Rei{\ss}, and M. Schwamb,
  accepted for publication in Eur. Phys. J. {\bf A} (2005);
  preprint nucl-th/0503033.

\bibitem{blomqvist} I. Blomqvist and J.M. Laget,
  Nucl. Phys. {\bf A280} 405 (1977).

\bibitem{DrH99} D. Drechsel, O. Hanstein, S. S. Kamalov, and L. Tiator,
 Nucl. Phys. {\bf A645} 145 (1999).

\bibitem{PoS87} H. P\"opping, P. U. Sauer, and X.-Z. Zang, Nucl. Phys.
 {\bf A474}, 557 (1987).

\bibitem{Yam} Y. Yamaguchi, 
  Phys. Rev. {\bf 95} 1628 (1954).  

\bibitem{Dress} E.T. Dressler, W.M. MacDonald, and J.S. O'Connell, 
  Phys. Rev. {\bf C20} 267 (1979).

\bibitem {schmidt96} R. Schmidt, H. Arenh{\"o}vel, and P. Wilhelm,
  Z. Phys. A {\bf 355} 421 (1996).

\bibitem{DaA03} E. M. Darwish, H. Arenh\"ovel, and M. Schwamb.
  Eur. Phys. J. {\bf A16} 111 (2003).

\end{thebibliography}
\end{document}